\documentclass[conference, a4paper]{IEEEtran}
\usepackage{cite}
\usepackage{hyperref}
\usepackage{amsmath,amssymb,amsfonts}
\usepackage{algorithmic}
\usepackage{graphicx}
\usepackage{textcomp}
\usepackage{xcolor}
\def\BibTeX{{\rm B\kern-.05em{\sc i\kern-.025em b}\kern-.08em
    T\kern-.1667em\lower.7ex\hbox{E}\kern-.125emX}}
\begin{document}

\title{Two-Tier UAV-based Low Power Wide Area Networks: A Testbed and  Experimentation Study}

\author{
\IEEEauthorblockN{Srdjan Sobot\IEEEauthorrefmark{1}, Milan Lukic\IEEEauthorrefmark{1}, Dusan Bortnik\IEEEauthorrefmark{1}, Vladimir Nikic\IEEEauthorrefmark{1}, Brena Lima\IEEEauthorrefmark{3} Marko Beko\IEEEauthorrefmark{2}\IEEEauthorrefmark{3}, Dejan Vukobratovic\IEEEauthorrefmark{1}}
\IEEEauthorblockA{
\IEEEauthorrefmark{1}Faculty of Technical Sciences, University of Novi Sad, Serbia\\
\IEEEauthorrefmark{2}Instituto de Telecomunica\c{c}\~{o}es, Instituto Superior T\' ecnico, Universidade de Lisboa, Lisbon, Portugal\\
\IEEEauthorrefmark{3}COPELABS, Universidade Lus\'{o}fona de Humanidades e Tecnologias, Lisbon, Portugal
%Email: tijana.devaja@uns.ac.rs, marko.beko@tecnico.ulisboa.pt, dejanv@uns.ac.rs
}}

\maketitle

\begin{abstract}
In this paper, we propose, design, deploy and demonstrate a two-tier Low Power Wide Area Network (LP WAN) system based on Unmanned Aerial Vehicle (UAV) base stations suitable for dynamic deployment in deep rural environments. The proposed UAV-based LP WAN network augments the existing macro-cellular LP WAN network (Tier 1) with an additional layer of mobile base stations (Tier 2) based also on LP WAN technology. Mobile Tier 2 LP WAN base stations provide connectivity to static or mobile LP WAN user equipment deployed in the areas without direct Tier 1 LP WAN network coverage. The proposed two-tier LP WAN network scenario is suitable for various agricultural, forestry and environmental applications such as livestock or wild animal monitoring. In this experimental work, we report the prototype that was successfully deployed and used in a real-world deep rural environment without Tier 1 LP WAN network coverage. 
\end{abstract}


\IEEEpeerreviewmaketitle

\section{Introduction}
\label{intro}

We start this paper with a motivating rural Internet of Things (IoT) example. Consider a scenario where a drone equipped with a camera monitors livestock in a remote mountainous area. Let us assume that a drone needs to periodically observe a herd (e.g., by taking an image), count the number of animals in the herd, and report this single number to a remote server. The main problem is the lack of cellular connectivity in mountainous areas, where many valleys or canyons represent coverage holes for the cellular network. To establish a drone-to-network connectivity, additional deployment of mobile UAV-based base stations that serve as relays covering the macro-cellular coverage holes is needed. We note that, in scenarios such as counting objects, a small amount of data is transferred from drones operating in a certain area to the UAV-based base station, and further to a rural macro-cellular base station. For this reason, we consider a two-tier LP WAN system, where a macro-cellular base station is augmented with UAV-based base stations maintaining their own LP WAN cells. In other words, data from IoT end devices is transmitted to the UAV-based Tier 2 LP WAN base station that, in the backhaul, act as an LP WAN device itself and relays the received data to the macro-cellular Tier 1 base station. In this paper, 3GPP NB-IoT is used as the Tier 1 macro-cellular LP WAN, while LoRa is used as the Tier 2 LP WAN.

Deployment of UAVs as a capacity/coverage extension of the cellular network infrastructure is gaining traction, not only in academic research \cite{Mozzaffari_2017}, but also in 3GPP Release 17 standardization work \cite{3GPP-UAV}. Apart from more general investigations on UAV-assisted IoT networks, covering topics such as UAV placement \cite{Liu_2020} and path optimization \cite{Moataz_2019}, current work on UAV-assisted LP WAN networks is limited. For the case of UAV-based NB-IoT networks, recent work in \cite{Mignardi_2021} provides a simulation-based performance analysis. Similar simulation-based study is presented in \cite{Castellanos_2020} focusing on a UAV-assisted agricultural IoT system deployed in a rural area. Finally, the case where NB-IoT/LoRa user equipment (UE) is mounted on a UAV and connected to a Tier 1 NB-IoT eNodeB (eNB) or LoRa gateway, which represents only the Tier 1 network in our scenario, is recently experimentally explored in \cite{Dambal_2019, Wang_2021}. 

In this paper, our goal is to present a testbed platform developed to support UAV-based LP WAN experimentation in deep rural environments outside of the cellular network coverage\footnote{This paper presents an extended version of the solution presented by a joint University of Novi Sad and Lusofona University team at the IEEE Vehicular Technology Society UAV Innovation Challenge \cite{VTS-UAV-Challenge}. The team won the first prize at both the first and the second (final) competition stage.}. Our goal is to share insights from our hands-on experience collected through the design, development, implementation and testing of a two-tier UAV-based NB-IoT/LoRa network. We use custom-designed NB-IoT nodes from which we are able to collect a wealth of data, e.g., radio channel conditions, fine-grained protocol messages exchanged between the UE and the eNB, NB-IoT module energy consumption, etc. The collected data, only a sample of which is presented in this initial report, will allow optimization of LP WAN network parameters, such as the size of the Tier 2 cell, position and altitude of the Tier 2 drone base station, and the required capacity of the NB-IoT backhaul link.

The paper is organized as follows. In Sec. II, we present the proposed two-tier LP-WAN system architecture. The experimental system implementation is described in detail in Sec. III. Real-world experiments and samples of collected data are illustrated in Sec. IV. Sec. V. concludes the paper.

\begin{figure*}[htpb]
\centering
\includegraphics[width=6.2in]{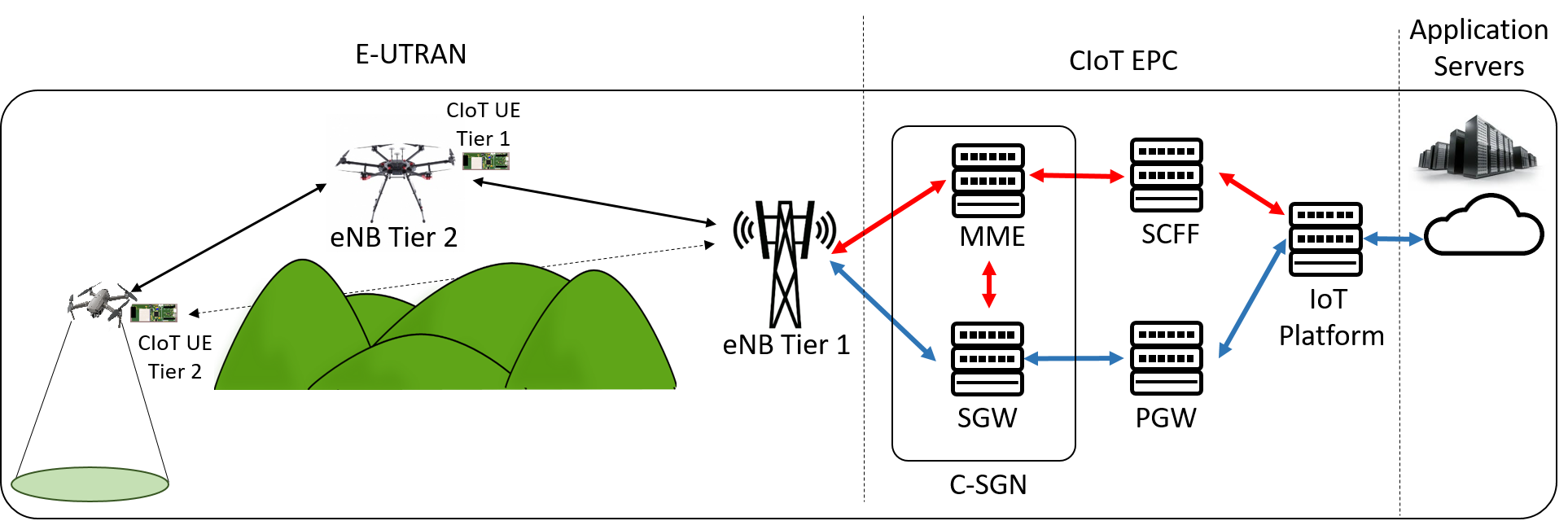}
\caption{Two-Tier Rural Cellular IoT Network Architecture.}
\label{Fig_1}
\end{figure*}

\section{Proposal for Two-Tier LP-WAN Solution}

\subsection{System Architecture}
\label{architecture}

Narrowband IoT (NB-IoT) is a new 3GPP Cellular Internet of Things (CIoT) technology that can be seamlessly integrated into an existing 3GPP 4G/5G architecture, coexisting with 4G LTE and 5G NR in the radio access network, while relying on the same evolved packet core (EPC) network \cite{book-nbiot}. Relevant 3GPP CIoT architecture elements are illustrated in Fig. \ref{Fig_1}. CIoT user equipment (CIoT UE) connects to the network via a neighbouring base station or eNB, a part of Evolved Universal Terrestrial Radio Access Network (E-UTRAN). NB-IoT downlink/uplink resources are allocated either within 4G LTE band (in-band deployment), at its edge (guard-band deployment), or as a separate channel (out-of-band deployment). After eNB, both user-plane and control-plane data is processed at CIoT Serving Gateway Node (C-SGN), which includes both control-plane Mobility Management Entity (MME) and user-plane Serving Gateway (SGW). User-plane data further flows through Packet Gateway (PGW) to the mobile operator IoT platform, from where the data is forwarded via Internet to external networks towards final IoT application servers. 

In our setup, we consider a mobile operator NB-IoT network as a Tier 1 network. NB-IoT-capable Tier 1 eNBs provide coverage to rural areas, however, in a mountainous terrain, large areas (canyons, valleys) stay out of reach of the Tier 1 NB-IoT network. To fill-up the coverage holes, we introduce a Tier 2 LP WAN network with UAV-mounted base stations (BS). For a backhaul link of the Tier 2 network, we use a Tier 1 NB-IoT link. As an alternative, one could use 4G LTE link as a backhaul towards the Tier 1 LTE network. However, NB-IoT backhauling is selected due to 20dB larger maximum coupling loss as compared to the LTE, which increases the operational radius of Tier 2 LP WAN network. In this work, as a solution for the Tier 2 LP WAN network, we use the Tier 2 LoRa network. We also include the demo of Tier 2 NB-IoT network albeit only in laboratory conditions.

In the scenario considered herein, Tier 2 LP WAN network user is a mobile device, more precisely, a drone or a group of drones, operating in the area of UAV-mounted Tier 2 LP WAN BS coverage. For this purpose, such a drone is equipped with an LP WAN module that is connected to the Tier 2 LP WAN network. Motivated by the example of livestock monitoring, any scenario where a drone operates in a remote area outside of cellular coverage, and where it generates a low amount of data (for example, a summary of specific computer vision tasks), falls into the proposed two-tier NB-IoT/LoRa network domain. Besides the counting applications, other examples include tracking (e.g., animal tracking), detection (e.g., smoke/fire detection, forest disease detection), and many others.   

%\subsection{System Elements}
%\label{elements}

% In this section, we present the details of the system elements.

\subsection{Tier 1 NB-IoT Network}

\textbf{Tier 1 NB-IoT eNB/EPC:} Our NB-IoT experimentation relies on an NB-IoT connectivity provided by a nation-wide mobile network provider A1 \cite{A1}. Thus in our demonstration, Tier 1 NB-IoT UE attached to the Tier 2 UAV BS will connect to the closest Tier 1 NB-IoT macro-cellular eNB in order to establish an NB-IoT-based backhaul connection for the UAV-mounted Tier 2 LP-WAN cell.   

\textbf{Tier 1 NB-IoT UE:} We use our own, custom-designed, NB-IoT UE devices (see Sec. 3 for more details). Tier 1 NB-IoT UE, mounted to the UAV, receives data from the Tier 2 LP WAN module, encapsulates it into a UDP packet, and initiates its transmission via Tier 1 NB-IoT network. More precisely, the firmware inside the on-board micro-controller unit (MCU) of the NB-IoT UE initiates network registration and subsequent communication tasks by issuing AT commands to the NB-IoT module.

\subsection{Tier 2 LoRa Network}

For the Tier 2 network implementation, we consider LoRa LP WAN technology that operates within unlicensed ISM bands (as opposed to the NB-IoT technology which utilizes licensed spectrum) \cite{LoRaWAN}. 
%The most widely used technology in this class is LoRaWAN, featuring communication range of about 15km in the outdoor environment. However, there are some drawbacks when it comes to the usage of LoRaWAN in our use case. It requires the infrastructure consisting of gateway(s) serving as access points, a network server and application server. 
LoRaWAN is an open standard, defining a MAC layer protocol, which utilizes LoRa modulation at the physical layer. To relieve the drone of carrying a power-hungry gateway, we use barebone LoRa modules based on Semtech SX1276/78 \cite{LoRa}. This way we strip down the wireless networking to the physical layer, using LoRa in the ad-hoc mode. Therefore, we use inexpensive low-weight low-power modules on both device and relay drone, at the expense of having to handle networking issues such as transmission scheduling, channel utilization, collision avoidance, integrity checking and message acknowledging through the firmware of the on-board MCU.

%Another reason to use LoRa network is due to unstable open-source software for emulating NB-IoT base station, but for consistency we tested NB-IoT Tear 2 eNB/EPC network under laboratory conditions. Also, we do not have permission from local authorities to emit NB-IoT signal in licensed ISM bands.

\textbf{Tier 2 Base Station:} In our experiments, as a Tier 2 BS, we use the DJI Matrice 600 Pro drone which has a total weight of 9.5kg and may carry an additional load of 6 kg. This drone it suitable for any load required to implement Tier 2 LoRa or NB-IoT base station. With a standard 6-pack of batteries and a maximum load of 6kg, the drone provides a hovering time of 16 mins, which increases by reducing the load to a maximum of 32 mins for an unloaded drone. 

\textbf{Tier 2 LoRa UE:} On the Tier 2 UE drone, we mounted a LoRa module based on Semtech SX1276/78 which is connected to the sensing platform that contains a set of sensors. As the Tier 2 UE drone, we use DJI Mavic, which is not designed to carry any additional load. For this reason, our Tier 2 UE is a lightweight device whose weight is less than 100g. 

%\subsection{Tier 2 NB-IoT Network (lab demo)}

%\textbf{Tier 2 NB-IoT eNB/EPC:} For the purpose of building an NB-IoT eNB, we use OpenAirInterface, which is an open and flexible 4G/5G experimentation platform created by the Mobile Communications Department at EURECOM \cite{Nikaein}. For LTE Core Network we use LTEBox, third-party EPC software developed by Nokia Bell Labs, and obtained from them under specific terms for academic use \cite{Chen}. For eNB transceiver functionality we use USRP software defined radio platform NI 2901 (B210) \cite{USRP}. This solution is tested only in lab conditions due to various constraints, among others, due to necessary permissions of emitting an NB-IoT signal in the licensed spectrum.
%Neglecting these constraints, we demonstrate that DJI Matrice 600 Pro drone can be used as an aerial eNB, containing Intel NUC Mini PC, USRP B210, NB-IoT module and an appropriate battery-pack to provide power supply for Mini-PC (and USB-attached USRP) as shown in Figure \ref{Fig_Tier1_drone}.

\section{Implementation of Two-Tier LP-WAN Solution}

In this section, we describe the solution that has been implemented and demonstrated. Before describing in detail the LP WAN scenarios that we have integrated and tested, we provide details on the equipment attached to each drone.

\subsection{Tier 2 (Small) UE Drone Configuration}

\begin{figure}[ht]
\centerline{\includegraphics[width=3.4in]{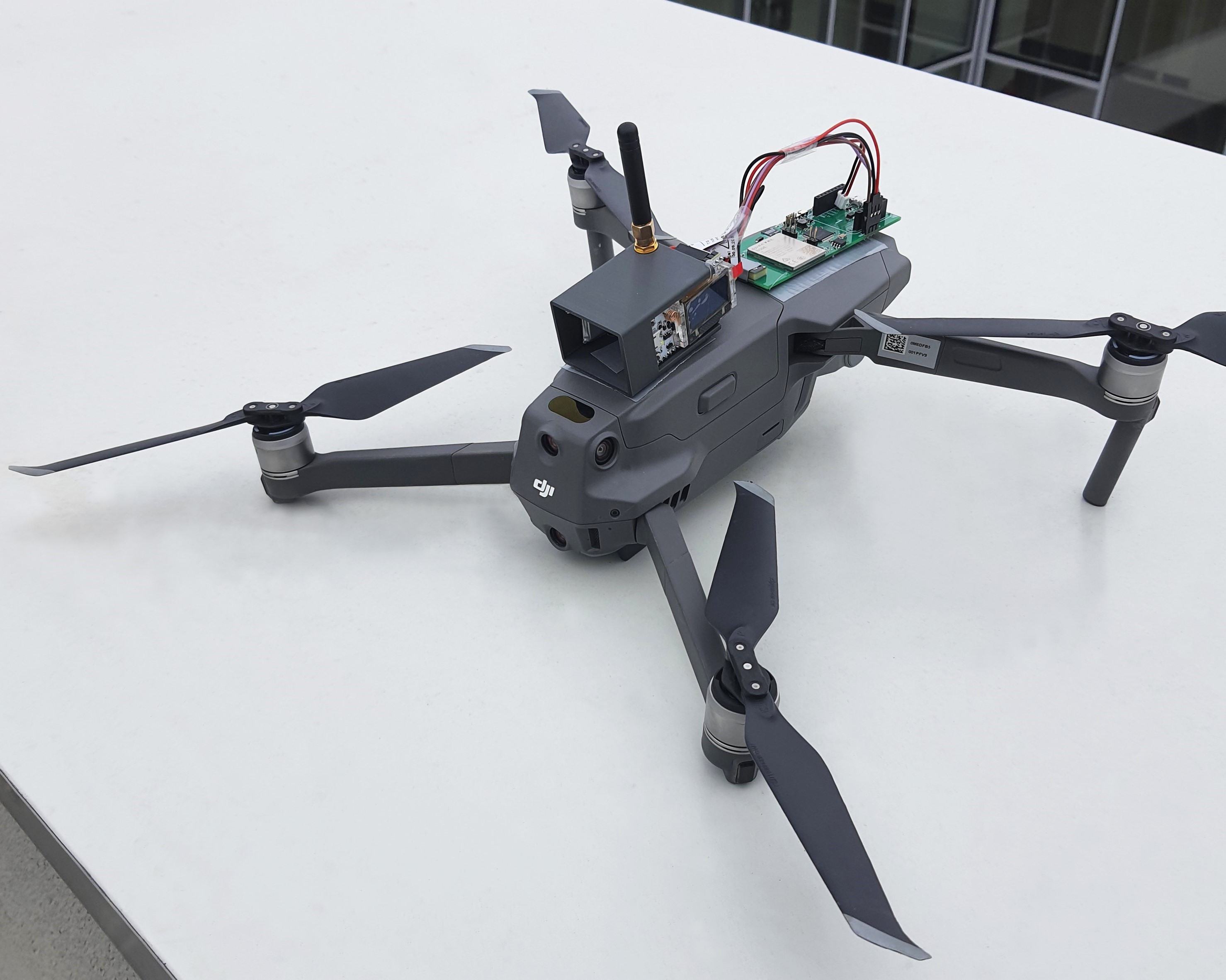}}
\caption{Tier 2 device drone with custom designed NB-IoT module, ESP32 platform with LoRa module and additional battery pack.}
\label{Fig_Tier2_drone}
\end{figure}

Figure \ref{Fig_Tier2_drone} illustrates the Tier 2 UE drone configuration representing a user device used in the experiment. We used DJI Mavic Pro 2 drone, due to its flat surface area at the upper side of the drone body, which is suitable to attach both the NB-IoT and LoRa end-user equipment. The DJI Mavic drone contains the following elements:

\textbf{LoRa node:} We use commercial off-the-shelf ESP32 platform with LoRa communication module \cite{ESP32_Lora} as a LoRa node.

Besides LoRa node, we also use an NB-IoT end-user equipment. The reason is twofold: i) to design experiments with Tier 2 NB-IoT network, and ii) to use it as a sensor platform to collect sensor data and send them via LoRa node in the experiment with Tier 2 LoRa network.

\begin{figure}[ht]
\centerline{\includegraphics[width=3.4in]{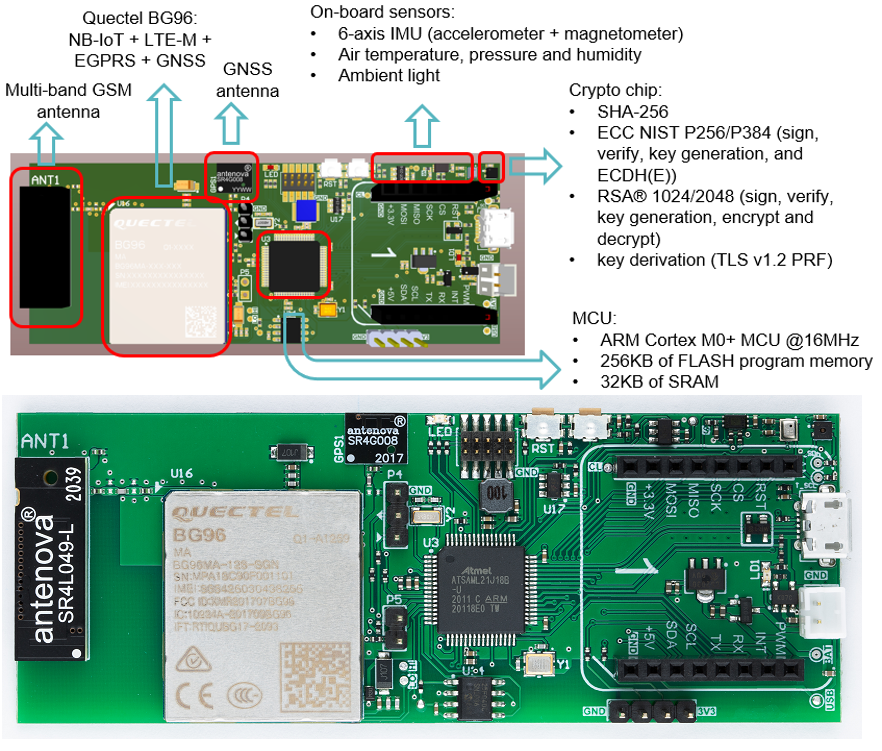}}
\caption{Custom-designed 3GPP NB-IoT/LTE-M node used in demonstration.}
\label{Fig_Edge_node}
\end{figure}

\textbf{NB-IoT node:} We use custom-designed NB-IoT platform (Figure \ref{Fig_Edge_node}) developed for experimentation, data collection and testing. The node integrates BG96 cellular IoT module from Quectel, which supports both NB-IoT and LTE-M. In addition, EGPRS is supported to ensure the connectivity in areas where LTE carrier might not be available. The integrated GNSS module provides the geo-location information. On-board sensors include the 6-axis Inertial Measurement Unit (IMU) that provides information about the vibrations and the magnetic field along X, Y and Z axes relative to the chip position. An additional set of sensors is used to measure the atmospheric conditions such as air temperature, pressure and humidity.

%In the Scenario 1 below (Tier 2 LoRa network), NB-IoT module is used only as a platform for sensing while ESP32 is used to transmit data via LoRa module, while in the Scenario 2 (Tier 2 NB-IoT network), NB-IoT module is used both for sensing and communication and ESP32 node is not used. 

\subsection{Tier 2 (Large) BS drone configuration}

\begin{figure}[ht]
\centerline{\includegraphics[width=3.4in]{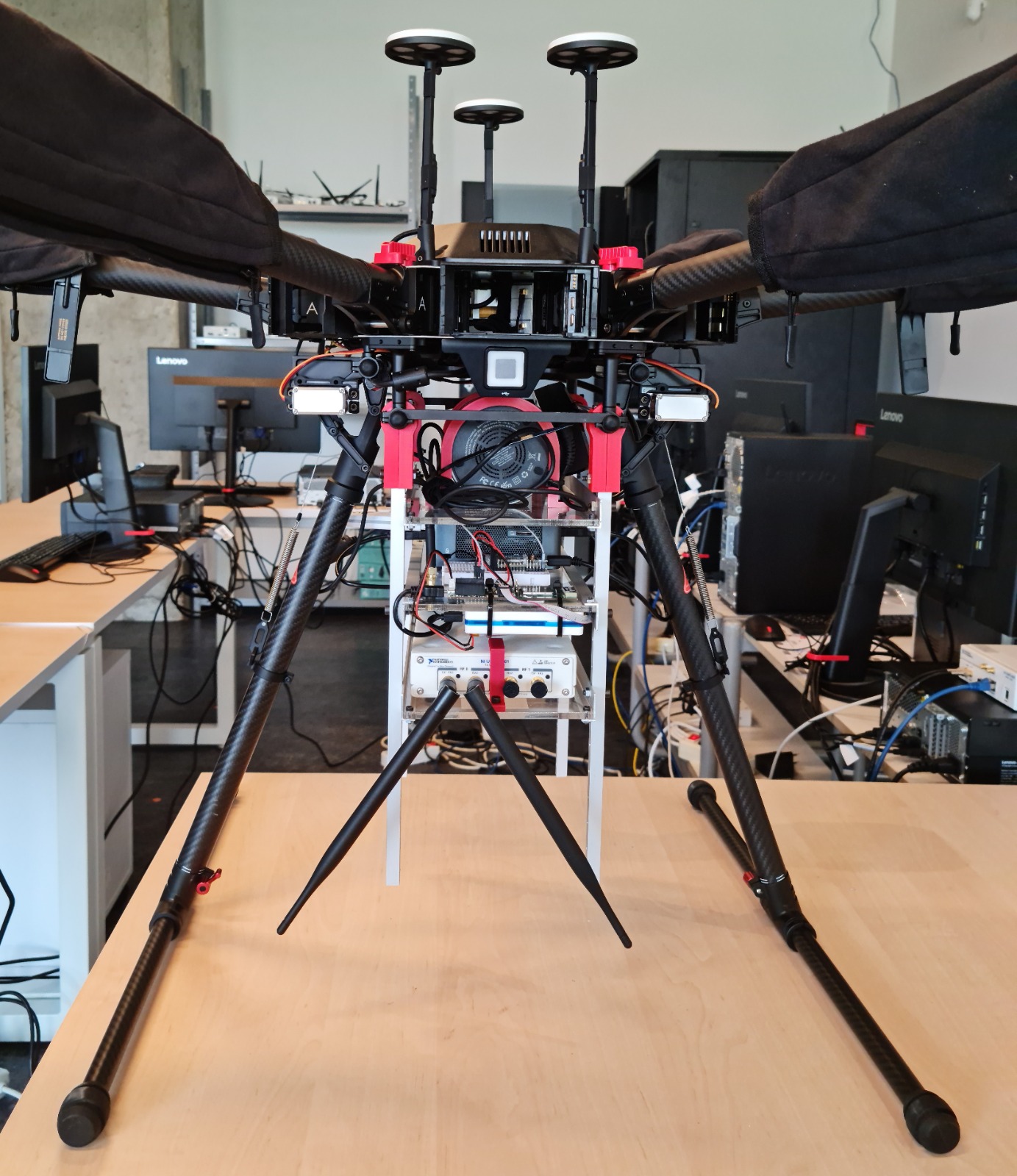}}
\caption{Tier 2 base station drone configuration with NB-IoT eNB, NB-IoT backhaul link, LoRa communication module and external battery pack.}
\label{Fig_Tier1_drone}
\end{figure}

Figure \ref{Fig_Tier1_drone} illustrates the large Tier 2 BS drone used in the experiment. The following elements are integrated on the custom-designed carriers attached to the DJI Matrice 600 Pro drone:
\begin{itemize}
    \item Intel NUC Intel i7 mini PC in combination with a software-defined radio USRP platform to serve as an open-source NB-IoT eNB. Mini PC runs OpenAirInterface NB-IoT software. 4G LTE Evolved Packet Core (EPC) software is also deployed at the mini PC.  
    \item Custom-designed NB-IoT module which serves as a backhaul link towards the mobile network infrastructure.
    \item Commercial off-the-shelf ESP32 platform with LoRa communication module.
    \item External battery pack to power up the equipment.
\end{itemize}

\subsection{Two-Tier LP WAN: Tier 1 NB-IoT + Tier 2 LoRa}

\begin{figure}[ht]
\centerline{\includegraphics[width=3.4in]{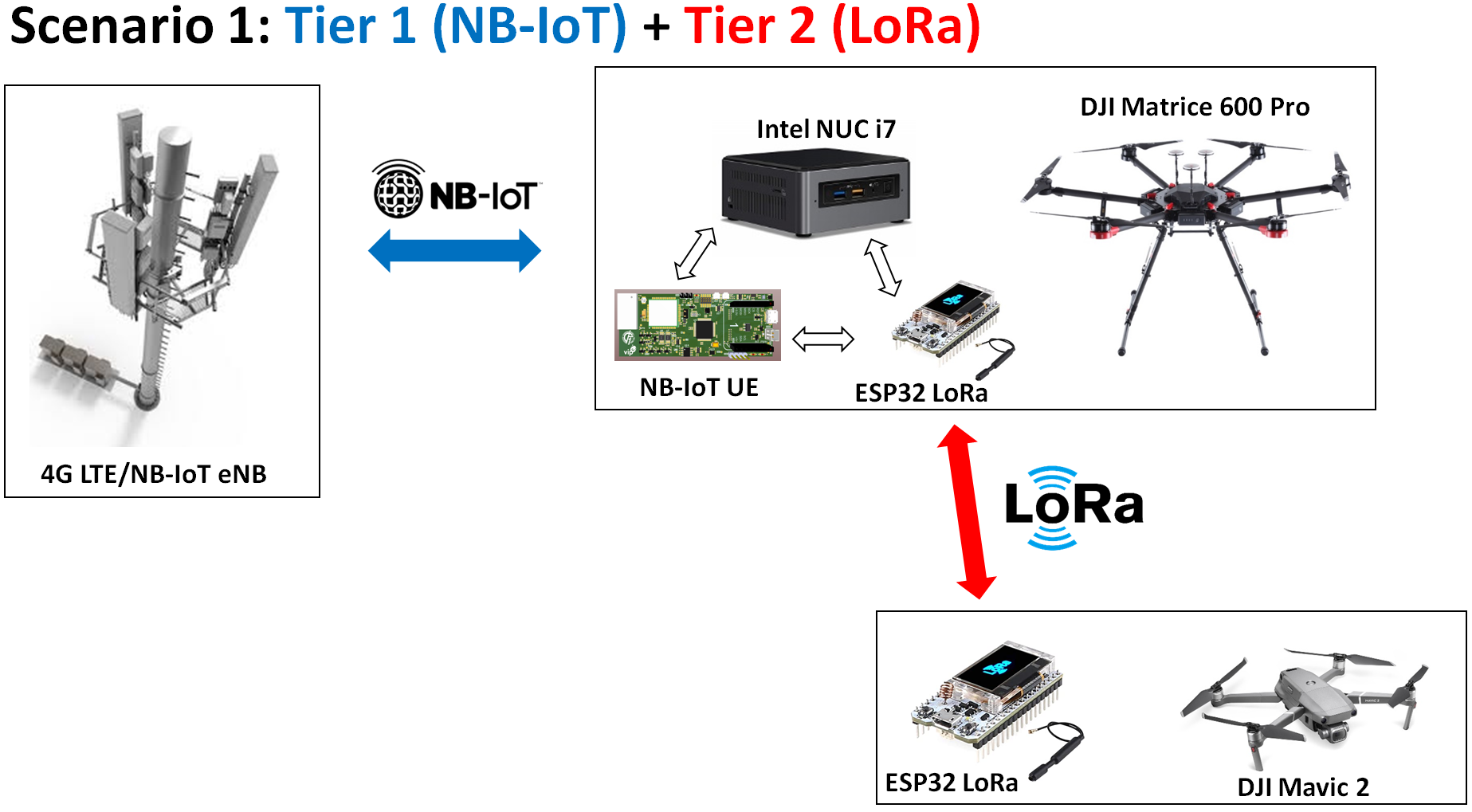}}
\caption{Scenario 1: Tier 1 NB-IoT and Tier 2 LoRa System.}
\label{Fig_VTS_scenario1}
\end{figure}

\begin{figure}[ht]
\centerline{\includegraphics[width=3.4in]{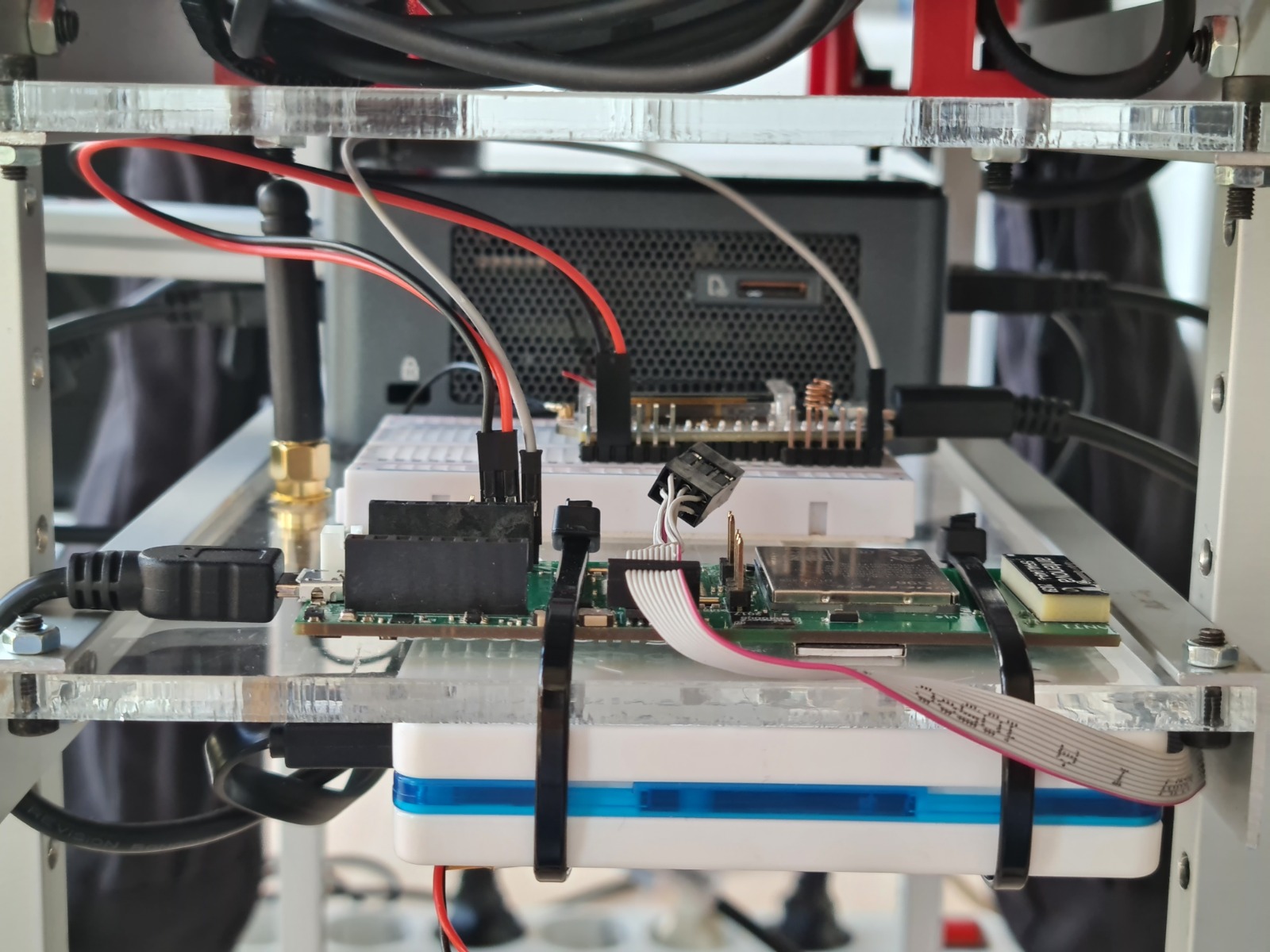}}
\caption{Scenario 1: Custom  designed  NB-IoT  module as backhaul link and ESP32 platform with LoRa communication module \cite{ESP32_Lora}.}
\label{Fig_Tier1_drone_scenario1}
\end{figure}

In Scenario 1: NB-IoT + LoRa, as shown in Figure \ref{Fig_VTS_scenario1}, communication between UAVs is established via LoRa link, while the backhaul connection from the Tier 2 BS drone to the Tier 1 eNB is established via the NB-IoT link. On the Tier 2 UE drone, information from sensors is transferred via LoRa module on ESP32 platform to the receiving LoRa device placed on the Tier 2 BS drone (Figure \ref{Fig_Tier1_drone_scenario1}). In addition, on the Tier 2 BS drone, we connected both the ESP32 platform and NB-IoT backhaul device to the Intel NUC. Using the screen-sharing app at Intel NUC and via a parallel 4G LTE connection (established through Intel NUC and USB-attached 4G LTE dongle), we monitored and logged information from both Tier 2 LoRa communication between the drones and Tier 1 NB-IoT backhaul communication towards Tier 1 macro-cellular NB-IoT eNB. In addition, information collected by sensors at Tier 2 UE drone and delivered via the proposed two-tier network in and presented (visualised) in the cloud-based application. This scenario is demonstrated in two environments: rural and urban, as detailed below. 

\subsubsection{NB-IoT+LoRa -- Rural Demo} 
\begin{figure}
\centerline{\includegraphics[width=3.3in]{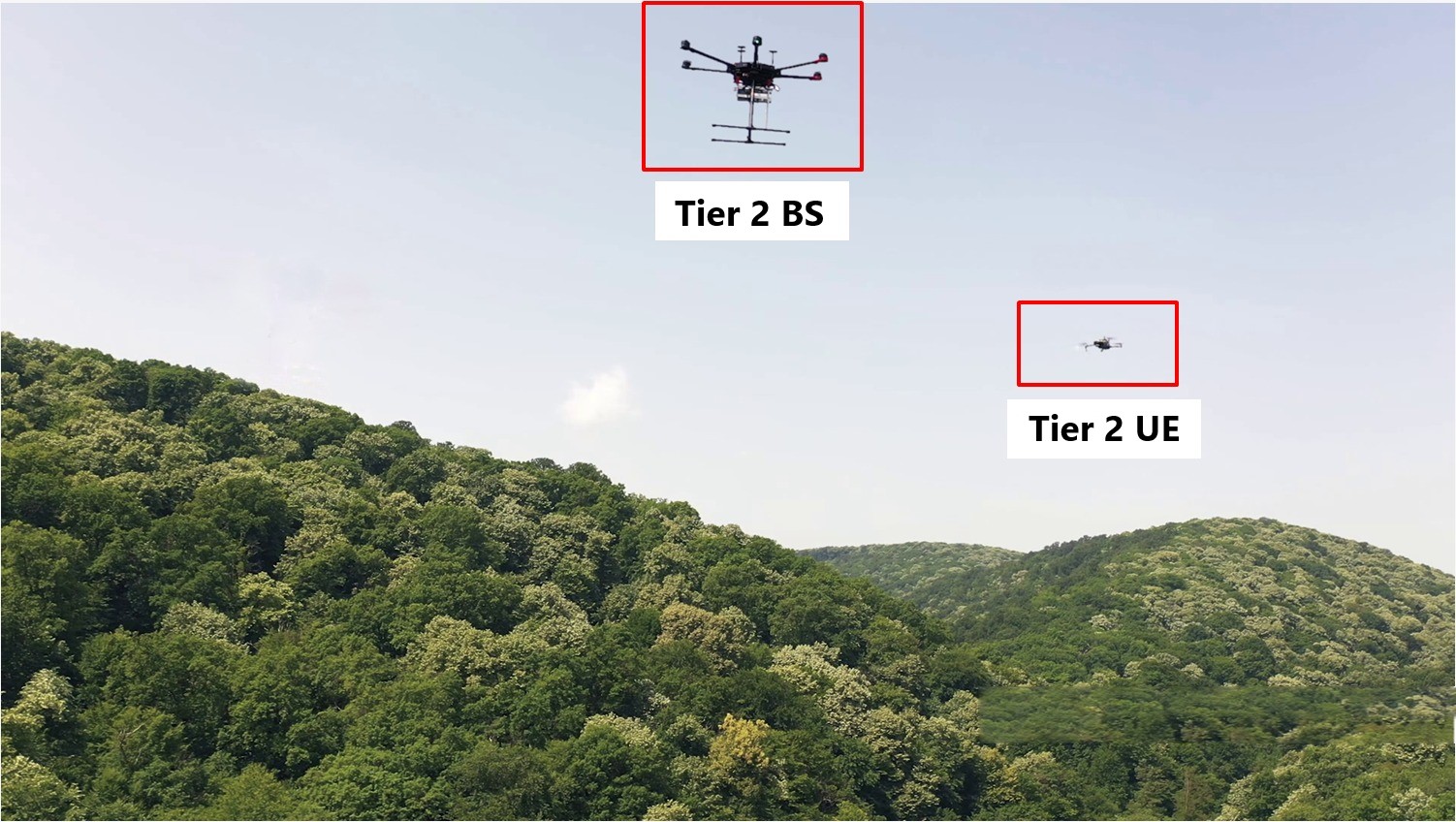}}
\caption{Real-world demonstration of Scenario 1 in rural setting.}
\label{Rural}
\end{figure}

\begin{figure}
\centerline{\includegraphics[width=3.3in]{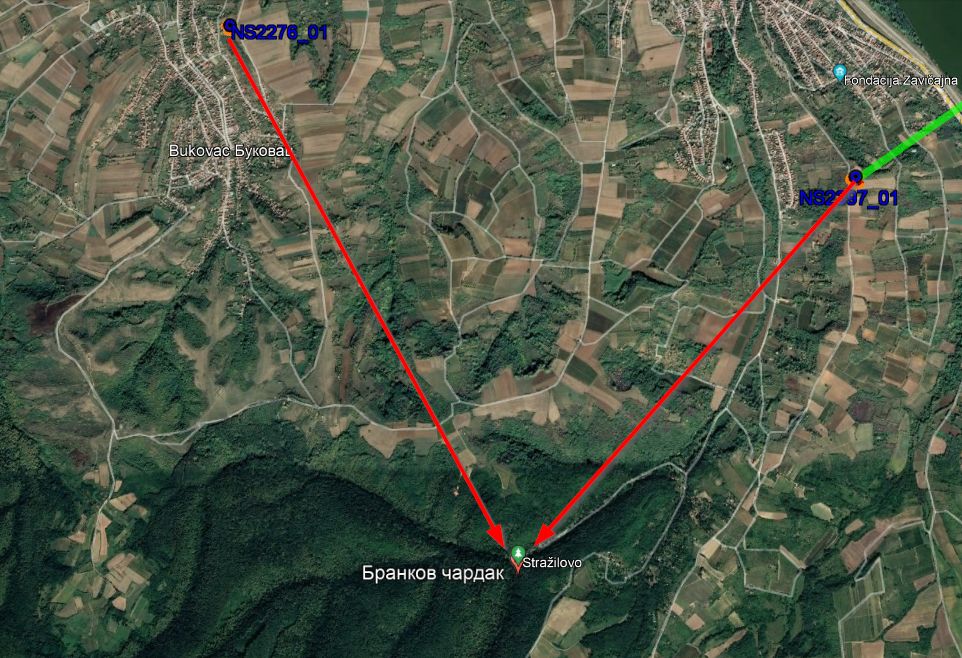}}
\caption{Coverage hole location -- Scenario 1 in rural setting.}
\label{BS-Strazilovo}
\end{figure}

The rural scenario demonstration took place in a valley of the Fruska Gora mountain (close to the city of Novi Sad, Serbia), as illustrated in Figure \ref{Rural}. Due to the terrain configuration, and position of the closest macro-cellular eNB sites, there was no service from the network operator at all (see Figure \ref{BS-Strazilovo}). We elevated the Tier 2 BS drone up to 70 meters from the ground, where the drone established a direct LoS with a nearby base station placed on the hill. This way, we created a LoRa cell below the Tier 2 BS drone covering the area of interest. Tier 2 UE drone was operating in the created cell, and it was able to send data from its on-board sensors to the Tier 2 BS drone and further via the NB-IoT backhaul link to the remote server application. Although end-to-end delay measurements are left for the future work, the data captured by Tier 2 UE drone were visualised at the server application in a near real-time fashion). Link to the full demonstration video is given at \cite{rural}.

\subsubsection{NB-IoT+LoRa  -- Urban Demo} 
\begin{figure}
\centerline{\includegraphics[width=3.2in]{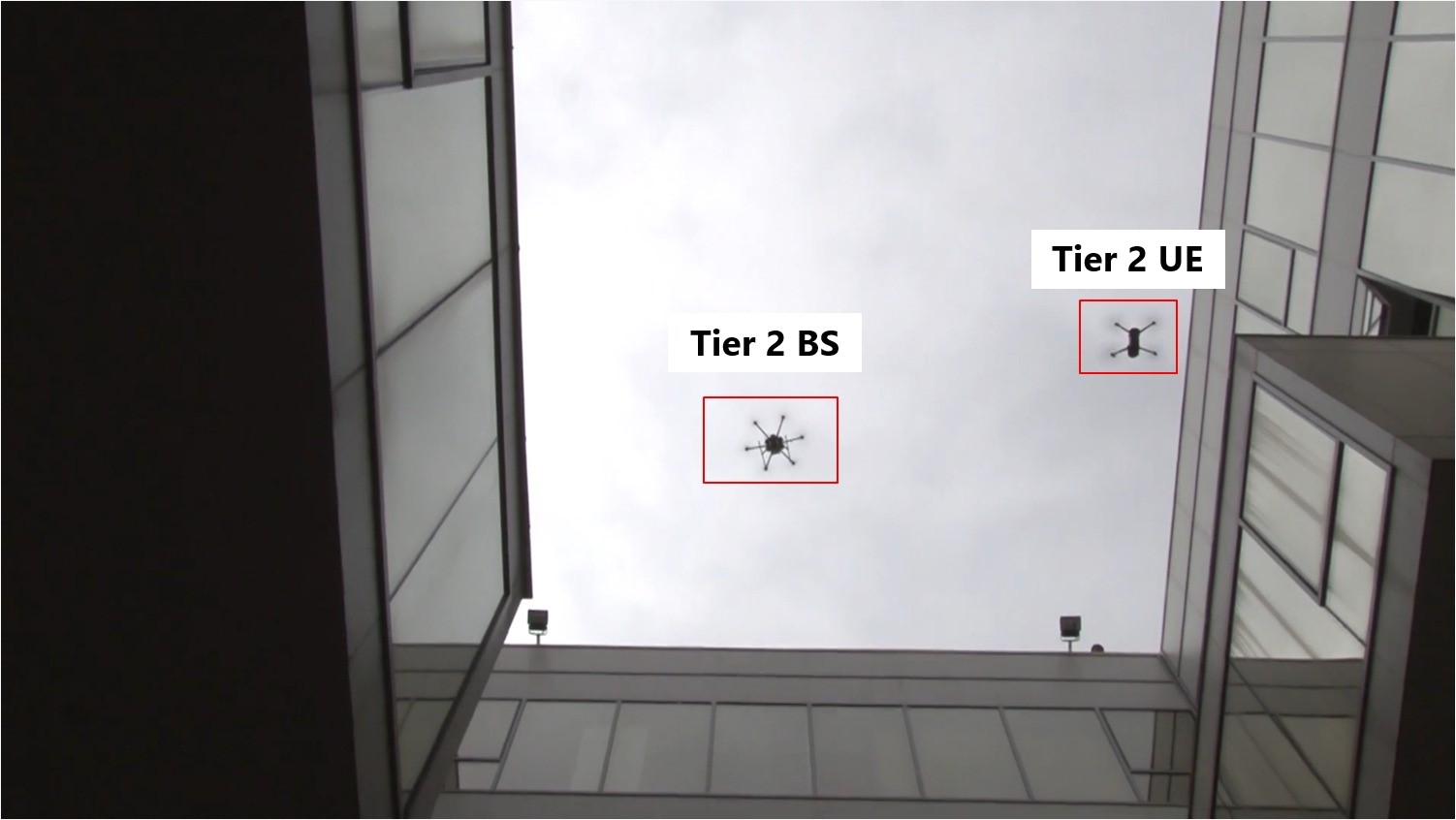}}
\caption{Real-world demonstration of Scenario 1 in urban setting.}
\label{Urban}
\end{figure}

To present our solution in an urban area, we took advantage of our Department building which has a large atrium in the central area, as shown in Figure \ref{Urban}. Due to a large reinforced glass facade of the building, its width and position of atrium, and  configurations of mobile operator Tier 1 eNBs around the building, the coverage is weak or does not exist in the building's atrium. Thus we placed Tier 2 UE drone inside the atrium, while Tier 2 BS drone was hovering above the building and had a LoS with a Tier 1 eNB. Similarly as before, the sensor data from the Tier 2 UE drone are transferred via the LoRa link to the Tier 2 BS drone, and further via the NB-IoT backhaul link to the remote server. Link to the full demonstration video is given at \cite{urban}.

%\begin{figure}
%\centerline{\includegraphics[width=3.2in]{Rural2.jpeg}}
%\caption{Rural scenario: LoRa cell created by Tier 2 base station UAV.}
%\label{Rural2}
%\end{figure}

\subsection{Two Tier LP WAN: Tier 1 NB-IoT and Tier 2 NB-IoT}

Besides Tier 2 LoRa network, we considered Tier 2 NB-IoT network based on OAI. Tier 2 BS is integrated on the large drone as an open-source OAI NB-IoT eNB along with a core network EPC software. Even though a ping request/response was successfully sent/received from the Tier 2 NB-IoT device, we were unable to successfully send a UDP packet from the Tier 2 NB-IoT device to the OAI NB-IoT base station at Tier 2 eNB drone (which is a known issue in OAI NB-IoT software which is still under development \cite{Chen}). However, the presented lab demonstration verified the viability of such a configuration.

%\begin{figure}[ht]
%\centerline{\includegraphics[width=3.2in]{VTS_scenatio2.jpeg}}
%\caption{Scenario 2: Tier 1 NB-IoT and Tier 2 OAI NB-IoT System.}
%\label{Fig_VTS_scenario2}
%\end{figure}

\begin{figure}
\centerline{\includegraphics[width=3.3in]{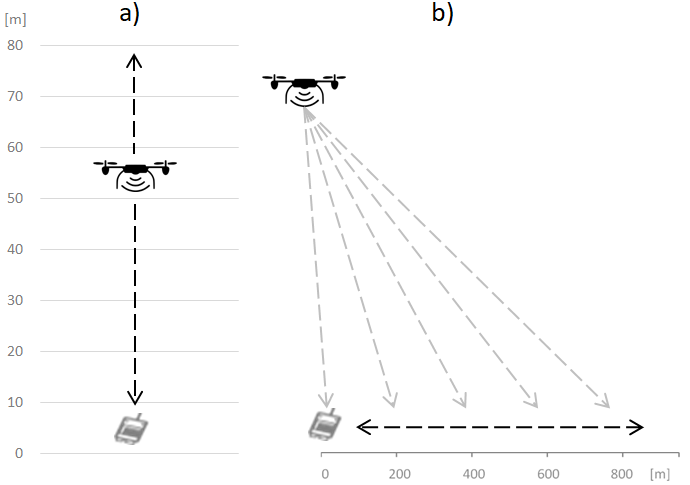}}
\caption{Rural scenario: a) Experiment I, b) Experiment II}
\label{Experiment}
\end{figure}

%\begin{figure}
%\centerline{\includegraphics[width=3.2in]{ExperimentI.png}}
%\caption{Rural scenario - NB-IoT backhaul link RSSI vs. drone height}
%\label{ExperimentI}
%\end{figure}

\section {Experimentation Results}

In this section, we present example results obtained from the real-world experiments described in Sec. III. In particular, the results are obtained in NB-IoT+LoRa case in both the rural and urban demo. We performed two experiments as depicted in Fig \ref{Experiment}. In Experiment I, the Tier 2 UE drone with LoRa device was stationary while the Tier 2 LoRa BS drone increased its altitude. In Experiment II, the Tier 2 LoRa BS drone hovered at a fixed height of 60m above the ground, while the Tier 2 LoRa UE drone moved close to the ground level in a straight path up to 800m away from the cell centre. In Experiment III, Tier 2 LoRa BS drone was changing altitude to record the quality of the backhaul Tier 1 NB-IoT link.

%\textbf{NB-IoT Backhaul Link Measurements (Experiment I):} Figure \ref{ExperimentI} shows radio signal measurements captured at the Tier 1 NB-IoT UE (mounted on a Tier 2 BS drone) of the signal delivered from the Tier 1 macro-cellular eNB. For different values of the drone height relative to the ground, the figure presents a variability of the Received Signal Strength Indicator (RSSI) values. At the height of up to about 20m from the ground, the Tier 1 NB-IoT device did not have a direct LoS with the Tier 1 eNB placed on a distant hill. Therefore, the signal was rather weak or there was no service at all. At 20 to 50 meters from the ground, the radio conditions become noticeably better. However, above 50m, the signal strength becomes noticeably improved due to a direct LoS connection with the Tier 1 eNB.

\begin{figure}
\centerline{\includegraphics[width=3.2in]{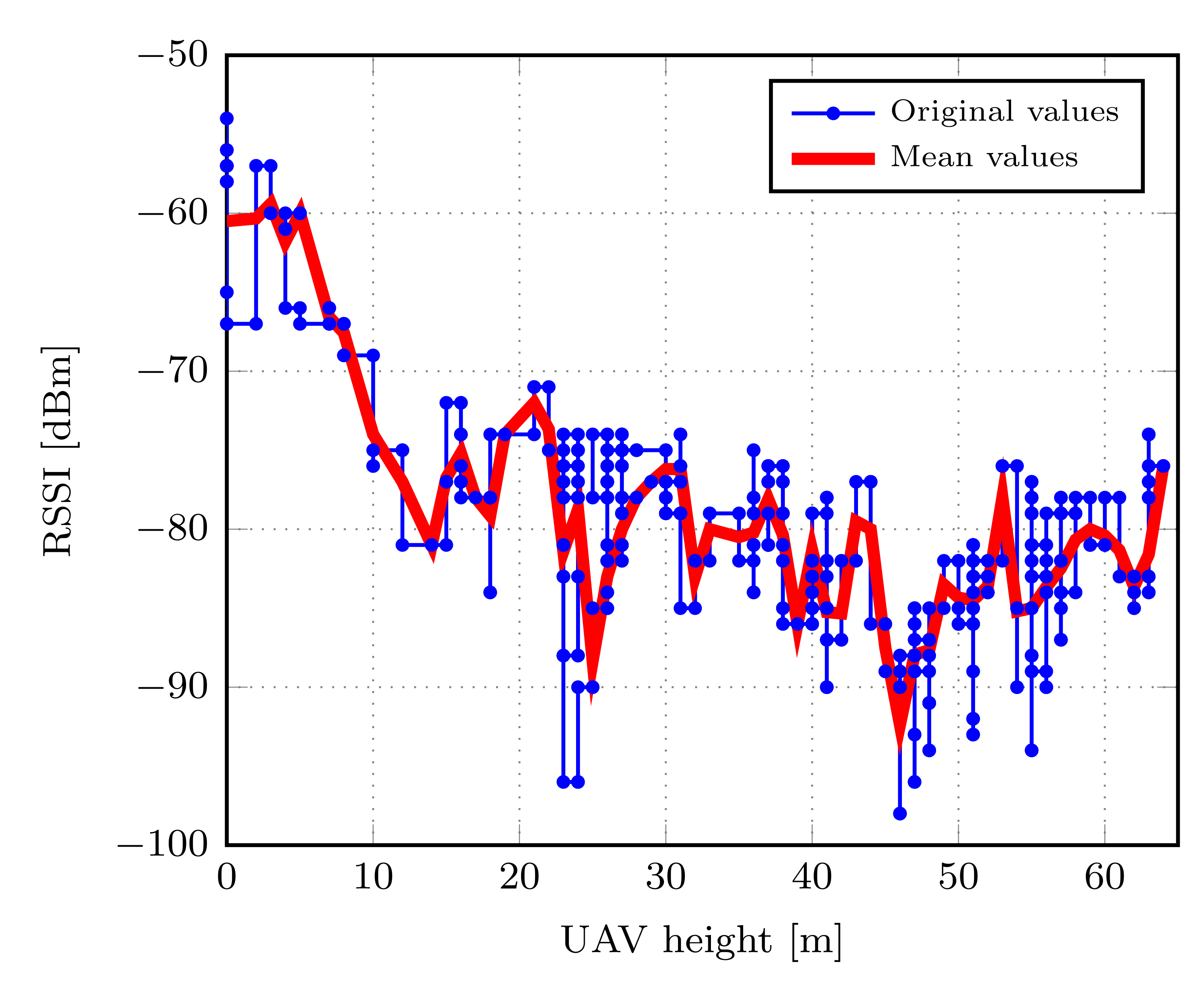}}
\caption{Rural scenario: Experiment I - LoRa RSSI vs. drone height}
\label{ExperimentI-LoRa}
\end{figure}

\textbf{Tier 2 LoRa Link Measurements (Experiment I):} Figure \ref{ExperimentI-LoRa} shows LoRa RSSI values captured at the Tier 2 LoRa receiver (on a Tier 2 BS drone) of the signal from a static Tier 2 LoRa UE drone as a function of the Tier 2 BS UAV height. The results confirm good radio conditions on a LoRa link, when the Tier 2 UE drone is almost at the ground level and exactly below the Tier 2 BS drone. 

\begin{figure}
\centerline{\includegraphics[width=3.2in]{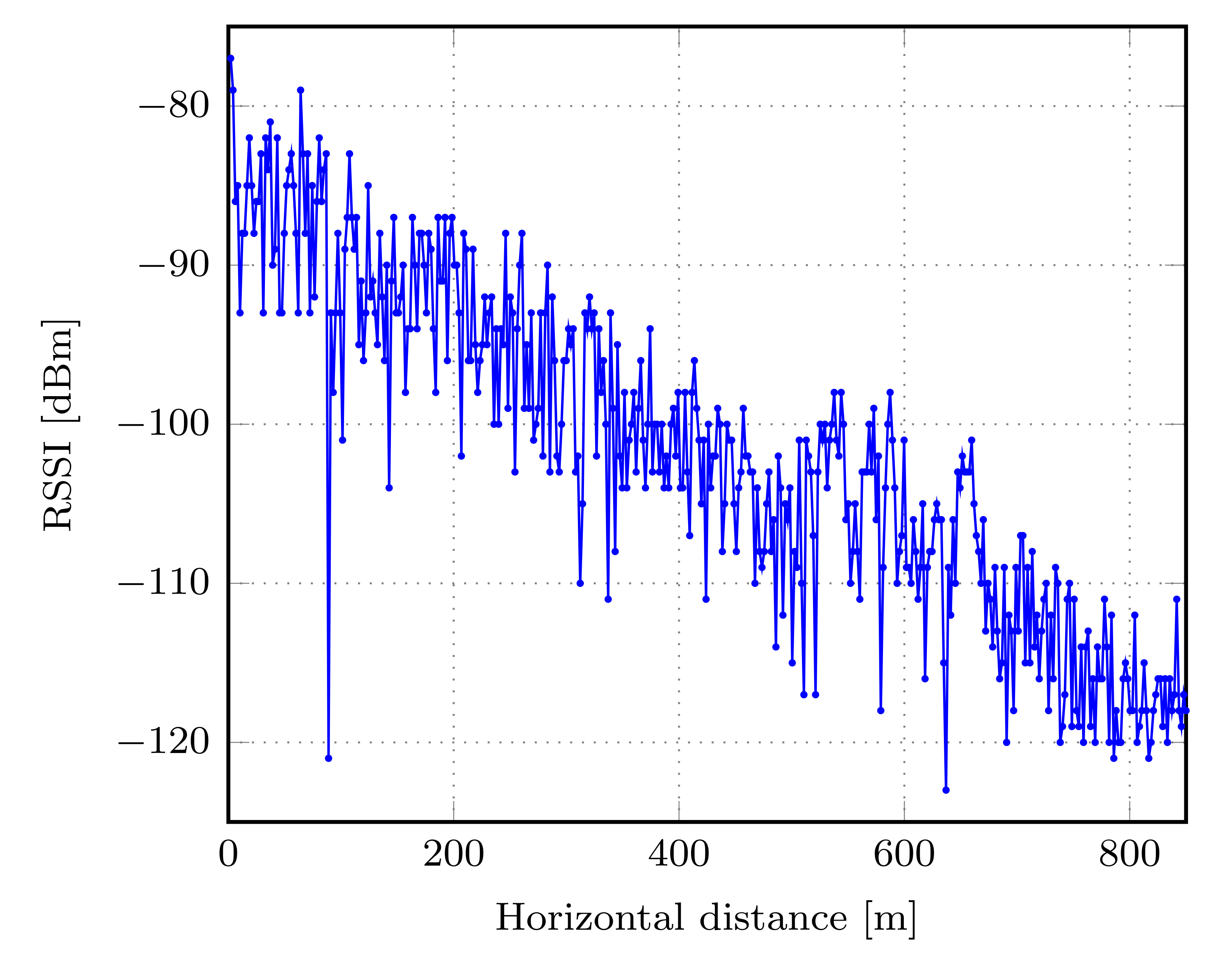}}
\caption{Rural scenario: Experiment II - LoRa RSSI vs. horizontal distance, h=60m}
\label{ExperimentII-LoRa}
\end{figure}

\textbf{Tier 2 LoRa Link Measurements (Experiment II):} Figure \ref{ExperimentII-LoRa} illustrates how RSSI values of the Tier 2 LoRa link decrease with horizontal distance of the Tier 2 UE drone from the centre of the Tier 2 LoRa cell. The Tier 2 UE drone was able to reach about 800m of horizontal distance from the cell centre before LoRa carrier is lost. Note that, in both experiments, the LoRa spreading factor has been set to the minimum value (SF7), which provides a smallest LoRa cell with the highest achievable data rate.

\begin{figure}
\centerline{\includegraphics[width=3.4in]{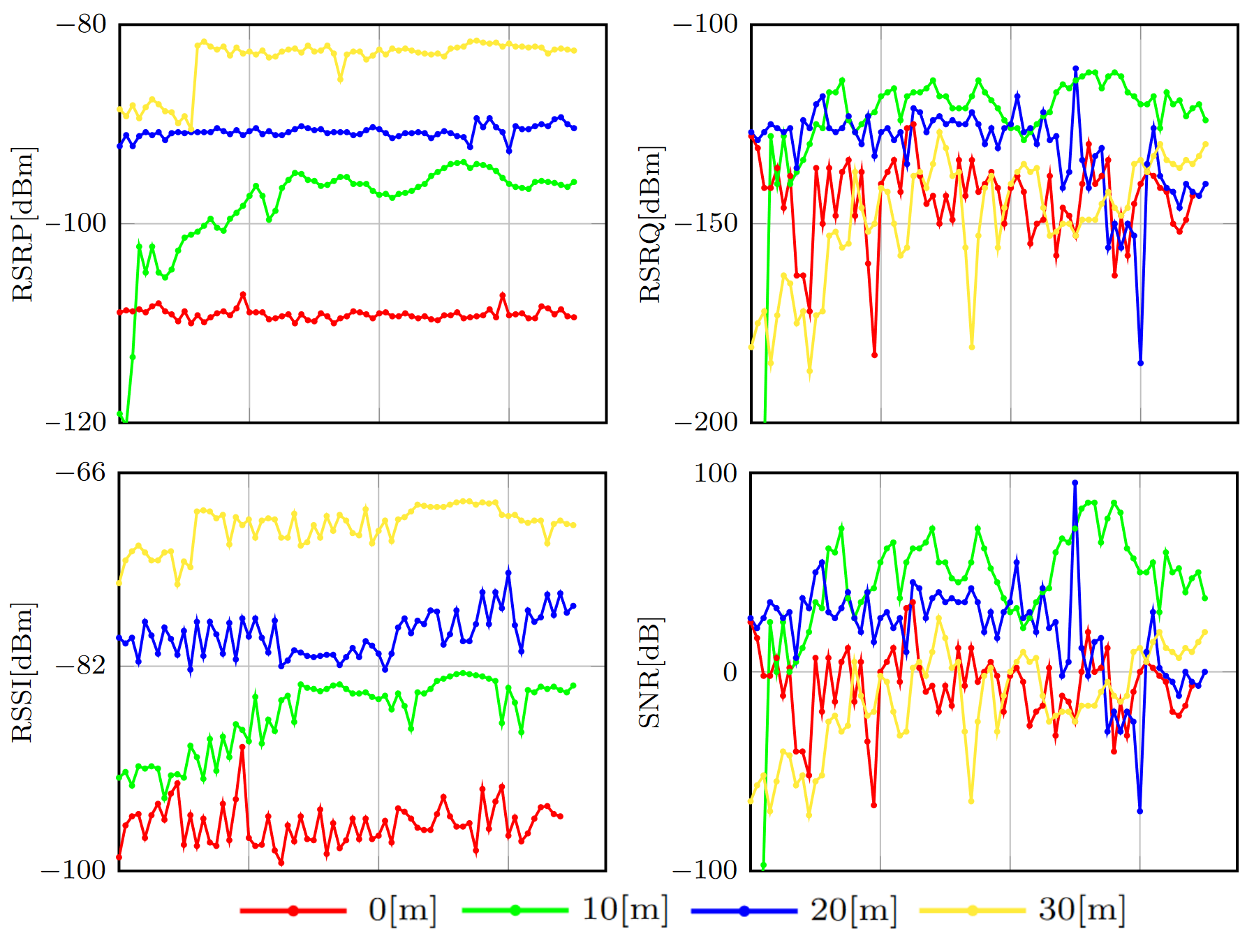}}
\caption{Urban scenario: Experiment III - NB-IoT RSRP, RSRQ, RSSI and SNR measurements for drone height h=\{0,10,20,30\}m}
\label{measurements}
\end{figure}

\textbf{Tier 1 NB-IoT Link Measurements (Experiment III):} Figure \ref{measurements} illustrates an example of logs of radio quality parameters (RSRP, RSRQ, RSSI and SNR) collected from the Tier 1 UE device placed at the large drone for different drone heights in urban scenario. These results can be used to evaluate the quality of the backhaul Tier 1 NB-IoT link.

\section{Conclusions}

In this paper, we proposed, designed, deployed and demonstrated a two-tier LP WAN architecture for deployment of IoT services in deep rural environments without a cellular network coverage. The proposed system is implemented in two scenarios (Tier 2 LP WAN being either LoRa or NB-IoT) and demonstrated in two real-world settings (urban and rural). The testbed and experimentation platform is presented through samples of radio link condition measurements, however, further data collection campaign is currently under way. Such measurement campaign will be used to verify analytical models for the optimal Tier 2 BS location and height selection.

\section{Acknowledgement}
This work was supported in part by the European Union’s Horizon 2020 Research and Innovation Programme under the Grant Agreement 856967, and in part by the Serbian Science Fund project AVIoTION.

% \section{Acknowledgment}


\begin{thebibliography}{1}

\bibitem{VTS-UAV-Challenge}
\href{https://events.vtsociety.org/vtc2020-fall/authors/vts-innovation-challenge-for-students/}{https://events.vtsociety.org/vtc2020-fall/authors/vts-innovation-challenge-for-students/}

\bibitem{Mozzaffari_2017}
M. Mozaffari, W. Saad, M. Bennis, M. Debbah, ``Mobile Unmanned Aerial Vehicles (UAVs) for Energy-Efficient Internet of Things Communications,'' \emph{IEEE Trans. Wirel. Commun.} Vol. 16, pp. 7574–-7589, 2017.

\bibitem{3GPP-UAV}
3GPP TS 22.125 V17.1.0. Unmanned Aerial System (UAS) Support in 3GPP. December 2019.

\bibitem{Liu_2020}
Y. Liu, K. Liu, J. Han, L. Zhu, Z. Xiao, and X.-G. Xia, ``Resource Allocation and 3-D Placement for UAV-Enabled Energy-Efficient IoT Communications,'' \emph{IEEE Internet of Things Journal,} Vol. 8, No. 3, pp. 1322--1333, 2020.

\bibitem{Moataz_2019}
S. Moataz, S. Sharafeddine, C. M. Assi, T. M. Nguyen, and A. Ghrayeb, ``UAV trajectory planning for data collection from time-constrained IoT devices,'' \emph{IEEE Trans. Wireless Comms.,} Vol. 19, No. 1, pp. 34--46, 2019.


\bibitem{Mignardi_2021}
D. Mignardi, R. Marini, R. Verdone, and C. Buratti, ``On the Performance of a UAV-Aided Wireless Network Based on NB-IoT," \emph{Drones}, Vol. 5, No. 3, 94: 2021.

\bibitem{Castellanos_2020}
G. Castellanos, M. Deruyck, L. Martens, and J. Wout, ``System assessment of WUSN using NB-IoT UAV-aided networks in potato crops,'' \emph{IEEE Access,} Vol. 8 (2020): 56823-56836.

\bibitem{Dambal_2019}
V.A. Dambal, S. Mohadikar, A. Kumbhar, I. Guvenc, ``Improving LoRa signal coverage in urban and sub-urban environments with UAVs,'' \emph{IEEE Int'l Workshop on Antenna Technology (iWAT),} pp. 210-213, 2019.

\bibitem{Wang_2021}
S.Y. Wang, J.E. Chang, H. Fan, Y.H. Sun, ``Comparing the Performance of NB-IoT, LTE Cat-M1, Sigfox, and LoRa for IoT End Devices Moving at High Speeds in the Air," \emph{Journ. of Signal Proc. Systems,} pp.1-19, 2021.

\bibitem{book-nbiot}
O. Liberg, M. Sundberg, et al., ``Cellular Internet of things: technologies, standards, and performance,'' Academic Press, 2017.

\bibitem{A1}
\href{https://a1.rs}{https://a1.rs}

\bibitem{LoRaWAN}
F. Adelantado, X. Vilajosana, P. Tuset-Peiro, B. Martinez, J. Melia-Segui, and T. Watteyne, ``Understanding the limits of LoRaWAN," \emph{IEEE Communications magazine,} 55(9), pp.34-40, 2017.

\bibitem{OAI}
\href{https://www.openairinterface.org/}{https://www.openairinterface.org/}

\bibitem{Nikaein}
Nikaein, N., Marina, M.K., Manickam, S., Dawson, A., Knopp, R. and Bonnet, C., 2014. OpenAirInterface: A flexible platform for 5G research. ACM SIGCOMM Computer Communication Review, 44(5), pp.33-38.

\bibitem{Chen}
Chen, C.C., Cheng, R.G., Ho, C.Y., Kanj, M., Mongazon-Cazavet, B. and Nikaein, N., 2020. Prototyping of Open Source NB-IoT Network. arXiv preprint arXiv:2006.02729.

\bibitem{USRP}
\href{https://www.ni.com/en-rs/support/model.usrp-2901.html}{https://www.ni.com/en-rs/support/model.usrp-2901.html}

\bibitem{LoRa}
\href{https://www.semtech.com/products/wireless-rf/lora-transceivers/sx1276}{https://www.semtech.com/products/wireless-rf/lora-transceivers/sx1276}

\bibitem{ESP32_Lora}
\href{https://heltec.org/project/wifi-lora-32/}{https://heltec.org/project/wifi-lora-32/}

\bibitem{urban}
\href{https://youtu.be/QU0vkbEcO1w}{https://youtu.be/QU0vkbEcO1w}

\bibitem{rural}
\href{https://www.youtube.com/watch?v=3_Px0hAjpVE}{https://www.youtube.com/watch?v=3\_Px0hAjpVE}

\end{thebibliography}
\end{document}